\documentclass{book}
\def\shorttitle{Importance of $0 \nu \beta \beta$ decay}
\usepackage{narosa}
\begin{document}

\markboth{Neutrinoless Double Beta Decay}{Importance of $0 \nu \beta \beta$ decay}
\def\pto{\mathrel{\vcenter{\hbox{${}~~{\cal P}$}\nointerlineskip\hbox{$\longleftrightarrow$}}}}
\def\oto{\mathrel{\vcenter{\hbox{${}~~{\cal O}$}\nointerlineskip\hbox{$\longrightarrow$}}}}
\def\cto{\mathrel{\vcenter{\hbox{${}~~~{\cal C}$}\nointerlineskip\hbox{$\longleftrightarrow$}}}}
\def\cpto{\mathrel{\vcenter{\hbox{${}~{\cal CP}$}\nointerlineskip\hbox{$\longleftrightarrow$}}}}
\def\C{{\cal C}}
\def\P{{\cal P}}
\def\T{{\cal T}}

\title{IMPORTANCE OF NEUTRINOLESS DOUBLE BETA DECAY }

\author{Utpal Sarkar}

\affil{Physical Research Laboratory, Ahmedabad - 380 009, INDIA}

\beginabstract

A natural explanation for the
smallness of the neutrino mass requires them to be Majorana particles
violating lepton number by two units. Since
lepton number violation can have several interesting
consequences in particle physics and cosmology, it is of utmost
importance to find out if there is lepton number violation in
nature and what is its magnitude. The neutrinoless double beta
decay experiment can answer these questions: if there is lepton
number violation and if neutrinos are Majorana particles.
In addition, the magnitude of neutrinoless double beta decay
will constrain any other lepton number violating processes.
This lepton number violation may also be relatd to the
matter-antimatter asymmetry of the universe, dark matter
and cosmological constant.

\endabstract

\section{Introduction}

Unlike any other particles, the existence of neutrinos was
postulated to explain energy-momentum conservation in the
beta decay. For our understanding of the macroscopic world,
the existence of neutrinos are not required. All the phenomena
we see around us can be explained by the electromagnetic and the
gravitational interactions. In addition the strong interaction
is required to explain how the positively charged protons could
stay together inside the nucleus. Only the beta decay involved
the neutrinos and it interacts very weakly.

The neutrinos are highly puzzling and it took more than sixty
years to find out that it has a very small mass. The atmospheric
and the solar neutrinos \cite{atm,sol}, combined with the
Laboratory experiments like KamLAND \cite{kam} have now established
that the mass-squared difference between any two of the three
neutrinos are non-vanishing:
\begin{eqnarray}
\Delta m_{atm}^2 &=& 2.1 \times 10^{-3}~{\rm eV}^2 ~~~~~{\rm with}
\sin^2 2\theta_{atm} > 0.92 \nonumber \\[.05in]
\Delta m_{sol}^2 &=& 7.9 \times 10^{-5}~{\rm eV}^2 ~~~~~{\rm with}
\tan^2 \theta_{sol} 0.4\pm 0.1  \,,
\end{eqnarray}
where $ \theta_{atm} $ is the mixing angle between $\nu_\mu$ and
$\nu_\tau$ and $ \theta_{sol}$ is the mixing angle between $\nu_e$
and one of the other two physical states, which is an admixture of
the states $\nu_\mu$ and $\nu_\tau$. The absolute mass of the neutrinos
have not yet been determined, although there is an upper bound
on the sum over all neutrino masses from cosmology \cite{wm}:
\begin{equation}
\sum_{i=e,\mu,\tau} ~m_{\nu_i} \leq ~ 0.69 ~{\rm eV} .
\end{equation}
The neutrinoless double beta decay also gives
an upper bound on the absolute mass of the neutrinos \cite{ndb},
but this bound is not valid if the neutrinos are Dirac particles.
We shall come back to this discussion later.

\section{Dirac and Majorana Neutrinos}

The smallness of the neutrino mass can be naturally explained in
the standard model, if the neutrinos are Majorana particles. A
Majorana particle has the property that it is its own antiparticle.
The main difference between a Majorana particle and a Dirac particle
lies in their mass terms:
\begin{eqnarray}
{\rm Majorana ~particle} ~~~&:&~~~M_{maj}~\Psi ~\Psi \nonumber \\
{\rm Dirac ~particle} ~~~&:&~~~M_{dir}~\bar \Psi ~\Psi . \nonumber
\end{eqnarray}
All charged fermions are Dirac particles, since the Majorana mass
terms do not conserve any charge. Only the neutrinos can have either
Dirac or Majorana masses. Since the neutrinos carry lepton number,
lepton number will be violated if neutrinos are Majorana particles.
In the standard model, lepton number is exactly conserved and we have
not observed any lepton number violation in nature so far.

Let us now write the mass terms in chiral notation. We define
the left-handed and right-handed particles as:
$$ \psi_L = {1  - \gamma_5 \over 2} ~ \psi
~~~~~{\rm and} ~~~~~ \psi_R = {1+\gamma_5 \over 2} \psi \,.$$
We now denote the parity transformation [$(\vec x,t) \leftrightarrow (-\vec x,t)$]
by $\P$, charge conjugation [particle $\leftrightarrow$ antiparticle]
by $\C$ and time reversal [$(\vec x,t) \leftrightarrow (\vec x,-t)$] as $\T$.
The chiral fields transform under $\C$, $\P$ and $\C\P$ as:
$$
\begin{array}{rcl@{\hspace{.5in}}rcl}
\psi_L &\pto& \psi_R & (\psi^c)_L &\pto & (\psi^c)_R \\[.05in]
\psi_L &\cto & {\psi^c}_L  & \psi_R &\cto & {\psi^c}_R \\[.05in]
\psi_L &\cpto & {\psi^c}_R  & \psi_R &\cpto& {\psi^c}_L \,,
\end{array} $$
where the charge conjugation is defined as $\psi^c = C \bar \psi^T =
C \gamma_0 \psi^*$, with $C = - i \gamma_2 \gamma_0$ so that
$(\psi^c)_L = (\psi_R)^c = {1 \over 2} ( 1 - \gamma_5) \psi^c$
and ${\psi^c}_R= (\psi_L)^c = {1 \over 2} ( 1 + \gamma_5) \psi^c$.
The $\C\P\T$ theorem ensures that the $\C\P$
conjugate states of any field must always be present.
So, any theory can have the left-handed fields $\psi_L$
and its $\C\P$ conjugate state ${\psi^c}_R$. The mass term requires
the field $\psi_R$  and its $\C\P$ conjugate state ${\psi^c}_L$.

Denoting a neutrino by $\psi$, the most general mass term can be
written as
\begin{eqnarray}
{\cal L}_{mass} &=&- {1 \over 2} m_L \overline{{\psi_L}^c} \psi_L
- {1 \over 2} m_R \overline{{\psi_R}^c} \psi_R
-m_D \bar \psi_R \psi_L + h.c. \nonumber \\[.05in]
&=& {1 \over 2} m_L \psi_L^T C^{-1} \psi_L
+ {1 \over 2} m_R \psi_R^T C^{-1} \psi_R
+ m_D \psi_L^T C^{-1} {\psi^c}_L + h.c. \nonumber \\[.05in]
&=& {1 \over 2} \pmatrix{\psi & \psi^c}^T_L C^{-1}
\pmatrix{m_L & m_D \cr m_D & m_R} \pmatrix{\psi \cr \psi^c}_L
={1 \over 2}  \Psi_L^T C^{-1} M \Psi_L ,
\end{eqnarray}
where $\Psi_L^T = \pmatrix{\psi & \psi^c}_L^T$ and $M$ is a $2
\times 2$ symmetric mass matrix $M = M^T $.

This general mass term contains most of the information required
for an understanding of the Dirac and Majorana masses of the
neutrinos. Any theory can have only the left-handed neutrinos and
its $\C\P$ conjugate state, but no right-handed neutrinos. In this
case the neutrino could be massless or can have a Majorana mass
$m_L$. When both the left-handed and right-handed neutrinos are present,
several possibilities can emerge.
\begin{itemize}
\item[$\bullet$] the neutrinos are massless, so there are two Weyl spinors.
\item[$\bullet$] $m_L=m_R=0$, so that the left-handed and the right-handed neutrinos
combine to form a Dirac neutrino.
\item[$\bullet$] $m_L = m_R \neq 0$ and $m_D=0$, so that there are two Majorana neutrinos
and the physical states are $\psi_L$ and $\psi_R$ with masses $m_L$ and $m_R$.
\item[$\bullet$] $m_L$ or $m_R$ or both are non-vanishing, and $m_D \neq 0$. In
this case also it corresponds to two Majorana neutrinos and the physical
states are admixtures of the states $\psi_L$ and $\psi_R$ with masses obtained
by diagonalising the mass matrix.
\end{itemize}
{\em The difference between a Majorana and a Dirac particle is
that, for a Dirac particle the mass term takes a left-handed particle ($\psi_L$)
to a right-handed particle ($\psi_R$), while for a Majorana particle the
mass term takes a left-handed particle ($\psi_L$) to a right-handed
antiparticle  (${\psi^c}_R$, which is a $\C\P$ conjugate of a left-handed particle)
or takes a right-handed particle ($\psi_R$) to a left-handed antiparticle
(${\psi^c}_L$). Another important difference between a Dirac and Majorana
particles is the conservation of charges. If neutrinos are Majorana particles,
then the mass term violates lepton number by two units}.

A direct consequence of the lepton number violation is neutrinoless
double beta decay.
In some even-even nuclei ordinary beta decay is forbidden, although
double beta decay (with and without two neutrinos)
could still be allowed
\begin{eqnarray}
n + n &\to& p + p + e^- + e^- + \nu_e + \nu_e \label{dbeq} \\
n + n &\to& p + p + e^- + e^- . \label{ndbeq}
\end{eqnarray}
The $2 \nu \beta \beta$ decay (equation \ref{dbeq}) has been observed,
in which the total kinetic energy of the
two electrons is less than the total energy available, while
for the neutrinoless double beta ($0 \nu \beta \beta$) decay
the total kinetic energy of the two
electrons is same as the $Q$ value of the decay. This makes it
possible to distinguish these two processes.

\begin{figure}[thb]
\vskip 0in \epsfxsize=50mm \centerline{\epsfbox{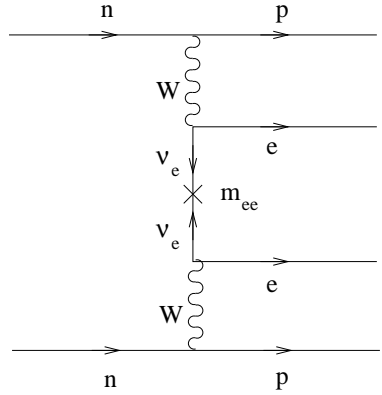}}
\vskip 0in \caption{Majorana mass of the neutrinos allowing
neutrinoless double beta decay. \label{ndbfg}}
\end{figure}

When the neutrinos are Majorana particles there will be total lepton
number $L$ violation, which will allow neutrinoless double beta decay
through the diagram given in figure \ref{ndbfg}. Here the neutrinos are
virtual particles in the intermediate state, so the neutrino masses and
mixing enter into the neutrino propagator. The half-life of the
neutrinoless double beta decay thus depends on the effective neutrino
mass that enters in the amplitude. The Heidelberg-Moscow $0 \nu \beta \beta$
decay experiment looked for the decay mode
$$ {}^{76}Ge \to {}^{76}Se ~ + ~ 2 ~e^-  $$
with their high resolution Ge detectors and given a strong bound
on the effective mass of the neutrinos \cite{ndb}, $m_{ee} < 0.2$ eV.
They also reported a few events for the $0 \nu \beta \beta$
decay, which is yet to be confirmed.

\section{Lepton Number Violation}

In the standard model, there are three left-handed neutrinos
$\nu_{iL}, ~i=e,\mu,\tau$ that transform under
$SU(3)_c \times SU(2)_L \times U(1)_Y$ as $(1,2,-1)$. Thus the Majorana
mass term is not allowed. Since there are no right-handed neutrinos,
the Dirac mass of the neutrinos are also not allowed. Thus neutrinos
are massless in the standard model. However, a natural explanation for
the observed tiny neutrino mass is possible in some extensions of
the standard model. A general approach to understand this is to
consider the most general dimension-5
effective lepton-number violating operator in the standard model
that can contribute to the Majorana masses of the neutrinos
\cite{op}
\begin{equation}
{\cal L}_{Maj} =  \Lambda^{-1} ( \nu \phi^\circ - e \phi^+ )^2 .
\label{nuop}
\end{equation}
Here $ \Lambda $ is some lepton-number violating heavy scale in
the theory and $\phi$ is the Higgs doublet scalar. The
electroweak symmetry breaking ($\langle \phi \rangle \neq 0$) then induce a
Majorana mass for the neutrinos
\begin{equation}
    {\cal L}_{Maj} = m_\nu \nu_{iL}^T ~C^{-1} \nu_{iL} ,
\end{equation}
where $m_\nu = v^2/\Lambda$. A large lepton number violating
scale $\Lambda$ can thus explain naturally why $m_\nu$ is much
smaller than the charged fermion masses. This also suggests that a
Majorana mass of the neutrinos is more natural than a Dirac mass.

The simplest extension of the standard model in which the effective
operator \ref{nuop} may be realized requires either right-handed
neutrinos or triplet Higgs scalars. In models with the right-handed
neutrinos, one extends the
standard model with three right-handed neutrinos $N_{i R}, i
= 1,2,3$, which are singlets under the standard model.
The mass terms for the neutrinos are now given by
\begin{eqnarray}
{\cal L}_{mass} &=& m_{D } ~ \nu_{L}
~{N^c}_L + M_R ~{N^c}_L ~ {N^c}_L + H.c. \nonumber  \\[.05in]
&=&
\stackrel{\pmatrix{ \nu & N^c}_L}{\phantom{x}}
 \pmatrix{0& m_{D}
\cr m^T_{D} & {M_R} }  \pmatrix{ \nu \cr N^c}_L .
\end{eqnarray}
Here the $3 \times 3$ mass matrix $m_D$ originates from the
standard model Higgs vacuum expectation
value, so it is of the order of charged lepton masses. But the
Majorana mass of the right-handed neutrinos, which is the lepton
number violating scale in the theory, could be very large:
$M_R \sim 10^{10}$ GeV. Thus
assuming $m_D \ll M$, the eigenvalues of this mass matrix then become,
\begin{equation}
m_1= -{m_D^2 \over M_R} ~~~~ {\rm and } ~~~~
m_2 = M_R  .
\end{equation}
We then get a light neutrino with mass $m_1 \sim 0.1$ eV, which is mostly
the left-handed neutrino with a small mixing
$\tan \theta = {2 M \over m_D} $ with the right-handed neutrino.
This is also known as the see-saw mechanism of neutrino masses \cite{seesaw}.
This small neutrino mass will contribute to the neutrinoless double
beta decay. Thus the neutrinoless double beta decay
can, in principle, determine the absolute mass scale of the neutrinos.

We shall now consider another equivalent realization of the effective
operator (\ref{nuop}), where the standard model is extended
to include a triplet Higgs scalar $\xi$, which
transforms under $ SU(3)_c \times SU(2)_L \times
U(1)_Y $ as $[1,3,+1]$ \cite{tripold,trip}.
Its couplings to the leptons and the standard model Higgs doublet $\phi$
break lepton number,
\begin{equation}
{\cal L}_{Yuk} = f_{ij} ~\xi ~ \ell_i \ell_j
+ \mu ~\xi^\dagger~ \phi \phi .
\end{equation}
We consider \cite{trip} the possibility $\mu \neq 0$
($\mu = 0$ models \cite{tripold} are ruled out by LEP data).
The neutral component of $\xi$ will acquire a induced $vev$ during the
electroweak symmetry breaking $u = { - \mu v^2 \over M^2} $,
where $M$ is the mass of the triplet Higgs $\xi$. The lepton
number is broken explicitly at a very high scale $M \sim \mu$.
So, there are no Goldstone bosons
corresponding to the broken lepton number symmetry.
The mass of the left-handed neutrinos are then given by
\begin{equation}
m^\nu_{ij} = f_{ij} u = - f_{ij} {  \mu v^2 \over M^2} ,
\end{equation}
which is of the order of $\sim$ eV.
The neutrino mass matrix is now directly proportional to the
Yukawa couplings $f_{ij}$. The absolute mass scale can be
determined by the neutrinoless double beta decay.

The smallness of the neutrino mass can thus be naturally
explained if neutrinos are Majorana particles and lepton number
is violated by two units. The Majorana nature of the neutrinos
can be confirmed by the neutrinoless double beta decay. In
fact, any lepton number violating processes can contribute to
the neutrinoless double beta decay. So, all lepton number violating
processes are constrained by the neutrinoless double beta decay
\cite{ndbcons}. In left-right symmetric models the right-handed
charged gauge boson mass and the right-handed neutrino mass could
be constrained by the present bound on the $0 \nu \beta \beta$ decay.
The inverse beta decay are also strongly constrained. The
leptoquarks, diquarks and other exotic scalar bilinears that couples
to two fermions of the standard model are also constrained by the
$0 \nu \beta \beta$ decay. In supersymmetric models all the
R-parity violating and lepton number violating couplings are
strongly constrained by the $0 \nu \beta \beta$ decay. Even in
R-parity conserving supersymmetric models, there could be lepton
number violation originating from the soft terms, which are also
constrained by the $0 \nu \beta \beta$ decay.
The compositeness scale for some models
with composite particles are also constrained by the $0 \nu \beta
\beta$ decay. Some of these constraints and the consequence of
$0 \nu \beta \beta$ decay in colliders will be reviewed in another
article by Prof. S.D. Rindani in this proceedings. We shall now
proceed to discuss some cosmological consequences of the lepton
number violation.

\section{Matter-Antimatter Asymmetry}

Our universe is composed mainly of matter and very little antimatter.
This matter dominance requires an explanation, since a natural choice
would be to start with a universe that is neutral with respect to
any conserved charges like baryon or lepton numbers. At present the
most popular explanation of this matter dominance in the universe
originates from the lepton number violation that is required for
the Majorana neutrino masses. This is known as leptogenesis.
The present limit on the amount of
lepton number violation coming from the neutrinoless
double beta decay is just enough to explain this matter dominance and
this predictability makes this scenario more appealing. To
establish this connection between the neutrinoless double
beta decay and leptogenesis, we shall discuss leptogenesis
in the see-saw model and the
triplet Higgs scalar model discussed in the previous section.

The generation of the baryon asymmetry of the universe requires
three ingredients: {\it i}) Baryon number ($B$) violation,
{\it ii}) $\C\P$ violation, and
{\it iii}) departure of the $B$-violating interactions from equilibrium.
On the other hand if lepton number ($L$) is violated satisfying
all the three conditions, then that will generate a
lepton asymmetry of the universe. In the standard model, both
$B$ and $L$ are global symmetries, but $(B+L)$ is broken by
quantum effects arising from anomalous triangle loop diagrams.
These anomaly induced $B+L$ violating processes are suppressed
by the quantum tunnelling probability. But at finite temperature,
during the period $10^2 < T < 10^{12}$ GeV,
these interactions become strong in the presence of
some static topological field configuration
called the sphalerons \cite{40}. As a result any existing $L$ asymmetry
of the universe will get converted to the required baryon
asymmetry of the universe, before the electroweak phase transition
\cite{38,39}.

In the see-saw mechanism of neutrino masses,
the Majorana masses of the heavy right-handed
neutrinos violate lepton number. $\C\P$ violation
comes from the complex Yukawa couplings and interference of tree
level decays with one-loop diagrams. These interactions can also
satisfy the out-of-equilibrium condition. Thus the decays of the
right-handed neutrinos into a lepton ($N_{Ri} \to \ell_{jL} + \bar \phi$)
and also an antilepton ($N_{Ri} \to {\ell_{jL}}^c + \phi$)
can generate a lepton asymmetry of the universe, which then
get converted to a baryon asymmetry of the universe in the presence
of the sphalerons.

The ${\cal CP}$ violation comes from an interference of the
tree level decays of the right-handed neutrinos and the one loop
diagrams:

\begin{figure}[thb]
\vskip 0in
\epsfxsize=115mm
\centerline{\epsfbox{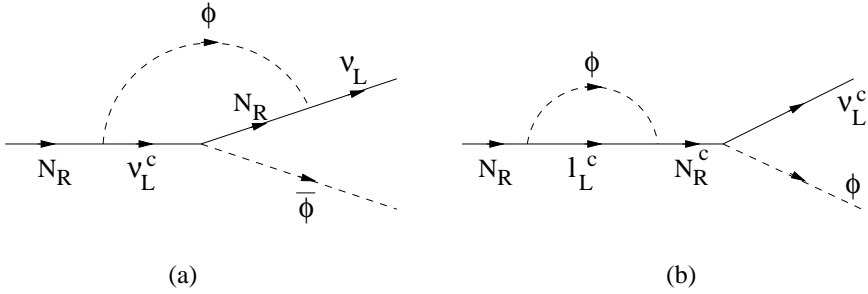}}
\vskip 0in
\caption{One loop (a) vertex and (b) self energy diagrams, which
interferes with the tree level right-handed neutrino decays to
produce CP violation.
\label{vert}}
\end{figure}

\begin{itemize}
\item[$(i)$] vertex diagram \cite{38,41} of figure \ref{vert}a,
which is similar to the $CP$ violation
coming from the penguin diagram in $K-$decays.

\item[$(ii)$] self energy diagram \cite{42} of figure \ref{vert}b,
which is
similar to the $CP$ violation in $K-\bar{K}$ oscillation, entering
in the mass matrix of the heavy Majorana neutrinos.

\end{itemize}

The interference gives an asymmetry
\begin{eqnarray}
\delta  &=& { \Gamma ( N \to \ell \phi^\dagger) -
\Gamma ( N \to \ell^c \phi) \over \Gamma ( N \to \ell \phi^\dagger) +
\Gamma ( N \to \ell^c \phi)}  , \label{lasy}
\end{eqnarray}
which, when satisfies the out-of-equilibrium condition:
\begin{equation}
\Gamma (N \to \ell \bar \phi) < 1.7 \sqrt{g_*} {T^2 \over M_P}
\hskip .5in {\rm at}~T = M_N,
\end{equation}
can generate the required lepton asymmetry of the universe.
Here the right-hand-side correspond to the expansion rate of the
universe and $M_P$ is the Planck scale. The lepton asymmetry thus
generated is same as the $(B-L)$ asymmetry of the universe, since there
is no primordial baryon asymmetry at this time. The sphaleron
interactions now convert this ${(B-L)}$ asymmetry to a baryon asymmetry of the
universe.

The amount of lepton asymmetry depends on the Yukawa couplings
and the out-of-equilibrium condition also depends on the Yukawa couplings.
Both these conditions can be satisfied for certain range of parameters,
which implies a neutrino mass of $m_\nu < 0.2$ eV \cite{buch}. Although this
limit is consistent with the upper bound on the neutrinoless double
beta decay, the reported events for the neutrinoless double beta decay
is not consistent with this limit \cite{ndb}. Thus determination of the
neutrinoless double beta decay half-life will tell us if the simplest version of
leptogenesis is possible.

\begin{figure}[thb]
\vskip -.3in
\epsfxsize=115mm
\centerline{\epsfbox{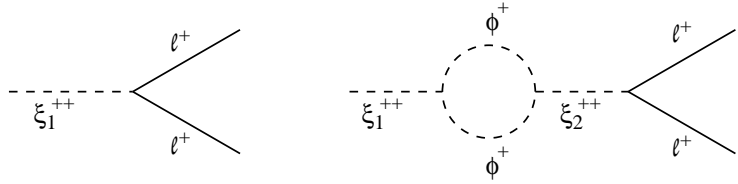}}
\vskip -.3in
\caption{The decay of $\xi_1^{++} \to l^+ l^+$ at tree level and in
one-loop order, whose interference gives ${\cal CP}$ violation.
\label{trl}}
\end{figure}

The triplet Higgs mechanism of neutrino masses \cite{trip}
can also allow leptogenesis. The decays of the triplet Higgs
scalars $\xi_a, a=1,2$, two
scalars are required for ${\cal CP}$ violation)
violate lepton number
\begin{equation}
\xi_a^{++} \rightarrow \left\{ \begin{array} {l@{\quad}l} l_i^+ l_j^+ &
(L = -2) \\ \phi^+ \phi^+ & (L = 0) \end{array} \right.
\end{equation}
$\C\P$ violation from the interference of the tree-level decays
and the self energy diagrams of figure \ref{trl}.
The rate of
$\xi_b \to \xi_a$ no longer remains to be the same as $\xi_b^* \to \xi_a^*$.
Since by $CPT$ theorem $\xi_b^* \to \xi_a^* \equiv \xi_a \to \xi_b$,
it means
\begin{equation}
\Gamma[\xi_a \to \xi_b] \neq \Gamma[\xi_b \to \xi_a] .
\end{equation}
This is a different kind of CP violation compared to the CP violation
in models with right-handed neutrinos. The
lepton asymmetry is now given by,
\begin{equation}
\delta = { \Gamma ( \xi \to \ell \ell) -
\Gamma ( \xi^\dagger \to \ell^c \ell^c) \over   \Gamma ( \xi \to \ell \ell) +
\Gamma ( \xi^\dagger \to \ell^c \ell^c) } .
\end{equation}
The out-of-equilibrium condition is satisfied when the triplet Higgs
scalars are very heavy. In this case the required amount of lepton
asymmetry do not constrain the neutrino masses.

\section{Dark Matter and Dark Energy}

The total matter in the universe is same as the critical density and
about 25\% of the matter is dark matter and about 70\% of matter is
in the form of dark energy. Only about 5\% matter is baryonic matter,
of which only a fraction is visible. There are several dark matter
candidates, including the lightest supersymmetric particle, which
is stable and very weakly interacting. One class of dark matter
candidate is related to the neutrino masses. If some discrete
symmetry forbids the Yukawa coupling relating the left-handed and
the right-handed neutrinos, there could be a second Higgs doublet
scalar that does not acquire any $vev$ or interact with the charged
fermions and remain inert. The lightest of these inert particles
(LIP) then could be a dark matter candidate \cite{dm}.

The problem with dark energy is that the large symmetry breaking scales in
particle physics would contribute orders of magnitude large
dark energy, while observations indicate that the dark energy
is comparable to the dark matter content of the universe. A
natural solution is thus to consider a scenario in which the
dark energy varies with time starting from a very high value
in the early universe. In a popular model, the mass density of a scalar field,
called the quintessence, gives the dark energy \cite{cc07a,cc07b}. The potential
of the quintessence field ensures that the decrease of the
dark energy is slower than the mass densities of matter and
radiation, so that the nucleosynthesis predictions are not
altered. Recently it has been pointed out that a varying neutrino
mass scenario can account for the dark energy of the universe \cite{mavans,mo}.
The variation of the neutrino mass can originate from some
scalar field, which could be a pseudo-Nambu-Goldstone boson
in the neutrino sector \cite{pngb}.

\end{document}